Superconducting Properties of adipic acid doped Bulk $MgB_2$ Superconductor

Arpita Vajpayee[1,2], V. P. S. Awana[1,*], G. L. Bhalla[2], A.K. Nigam[3] and H. Kishan[1]

[1]Superconductivity and Cryogenics Division, National Physical Laboratory, Dr K.S. Krishnan Road, New Delhi-110012, India

[2]Deparment of Physics and Astrophysics, University of Delhi, New Delhi-110007, India

[3]Tata Institute of Fundamental Research, Homi Bhabha Road, Mumbai 400005, India

**Abstract**

We report the effect of adipic acid ($C_6H_{10}O_4$) doping on lattice parameters, microstructure, critical temperature ($T_c$), current density ($J_c$), and irreversibility field ($H_{irr}$) for $MgB_2$ superconductor. Actual carbon (C) substitution level for boron (B) is estimated to be from 0.40 at% to 2.95 at% for different doping levels. A reduction in $T_c$ from 38.43 to 34.93 K and in lattice parameter 'a' from 3.084(3) Å to 3.075(6) Å is observed for the10 wt% $C_6H_{10}O_4$ doped sample in comparison to pristine $MgB_2$. This is an indication of C substitution at boron sites, with the C coming from the decomposition of $C_6H_{10}O_4$ at the time of reaction. Interestingly the doped samples have resulted in significant enhancement of $J_c$ and $H_{irr}$. All the doped samples exhibit the $J_c$ value of the order of $10^4$ A/cm$^2$ at 5 K and 8 T, which is higher by an order of magnitude as compared to undoped sample. This result indicates that $C_6H_{10}O_4$ is a promising material for $MgB_2$ for obtaining the excellent $J_c$ values under higher magnetic fields.

*: Corresponding Author
Dr. V.P.S. Awana
National Physical Laboratory, New Delhi-110012, India
Fax No. 0091-11-25626938: Phone no. 0091-11-25748709
e-mail: awana@mail.nplindia.ernet.in; www.freewebs.com/vpsawana/

## Introduction

Magnesium diboride ($MgB_2$) is a potential competitor to Nb-based superconductor due to its higher transition temperature ($T_c$), lower raw material cost and good upper critical field $H_{c2}$ & critical current density $J_c(H)$ values [1,2]. Further improvements in critical current density ($J_c$), upper critical field ($H_{c2}$) and irreversibility field ($H_{irr}$) is required for most of the practical applications because pure $MgB_2$ exhibit rather poor flux pinning behavior. An effective way to improve flux pinning is to introduce foreign pinning centers in $MgB_2$ because $MgB_2$ has large coherence length and small anisotropy. Therefore, chemical doping emerges as a simple and readily scalable technique to pin the fluxoids and thus improving the superconducting performance of pristine $MgB_2$ compound. Number of elements and compounds had been added previously to improve the superconducting performance parameters i.e., $J_c$, $H_{c2}$ and $H_{irr}$ of $MgB_2$ superconductor [3-13]. Among the various additives the C-containing compounds, such as SiC, C, $B_4C$ or carbon nanotubes (CNT), have been found quite effective dopants [14-17].

It is to be pointed out that the improvement of flux pinning depends on the size (*nano*) of particle doped in $MgB_2$. However use of *nano*-particles brings dilemmas such as higher cost and their agglomeration limiting the homogeneity of their mixing with $MgB_2$. For various forms of carbon doping, the substitution of boron cannot be achieved at the same temperatures as that of formation temperatures of $MgB_2$ due to their poor reactivity. In order to overcome these problems, some carbohydrates are proposed to be used as the dopants [18]. In this article, adipic acid ($C_6H_{10}O_4$) (aliphatic organic acid) is used as a dopant since it is a good candidate to be as a C source material for doping in $MgB_2$. The melting point of adipic acid is 152$^o$C i.e. it decomposes at temperature that is far below than the formation temperature of $MgB_2$ phase. Hence it produces fresh and highly reactive carbon on atomic scale at the time of the formation of $MgB_2$, which can substitute B easily than the other forms of C. The effect of adipic acid doping on lattice parameters (a and c), critical temperature ($T_c$), critical current density $J_c(H)$, irreversibility field ($H_{irr}$) , flux pinning force ($F_p$) and microstructures of pristine $MgB_2$ is presented here.

## Experimental Procedure

Polycrystalline $MgB_2$+(adipic acid)$_x$; (x = 0, 1%, 3%, 5%, 7% & 10%) samples were synthesized by solid-state reaction route. The constituent powders (Magnesium and Boron) were well mixed in stoichiometric ratio through grinding for 1.5 hour. The desired amount of adipic acid powder was dissolved in 10 ml of Acetone. Then we added this solution in ground $MgB_2$ raw powder, which was followed by second grinding in order to form the homogenous mixture. The mixture was pelletized using hydraulic press. The pellets were enclosed in soft iron tube and then annealed at 850$^o$C in argon atmosphere for 2.5 hours. The heating rate was about 425$^o$C per hour. After annealing, the system was allowed to cool down naturally. The optimization of synthesis parameters for the pure $MgB_2$ and its complete physical property characterization are discussed in detail in ref. 19. The X-ray diffraction (XRD) pattern of compounds was taken using $CuK_\alpha$ radiation. The resistivity $\rho(T)$ measurements were carried out using four-probe technique. The temperature is measured with an accuracy of ± 0.1K. The scanning electron microscopy (*SEM*) studies were carried out on prepared samples using a Leo 440 (Oxford Microscopy: UK) instrument. The magnetization measurements were carried out using *Quantum Design PPMS*, equipped with *VSM* attachment.

## Results and Discussion

X-ray diffraction pattern of the doped and undoped samples is shown in Fig. 1. It is found that undoped $MgB_2$ sample has well-developed hexagonal $MgB_2$ phase with a little amount of MgO (marked with * in figure). The presence of small quantity of MgO along with main phase of $MgB_2$ is consistent with earlier reports on similar samples [20,21]. Fitted and observed X-ray diffraction patterns of pure $MgB_2$ are shown in ref.19. The intensity of MgO peak increases monotonically with increasing amount of adipic acid. The peak situated between $2\theta = 33^o$ and $2\theta = 34^o$ shifts towards the higher $2\theta$ values with increasing x, indicating the contraction in a-axis in crystal lattice. The inset of Fig. 1 shows the shift of (100) peak confirming the decrease in ‘a’ parameter. The lattice parameters calculated from the XRD pattern show a large decrease in a-axis parameter but negligible change in c-axis parameter. As shown in Fig. 2(a), lattice parameter 'a' decreases from 3.084(3) Å for the undoped sample to 3.075(6) Å for the sample with the highest doping level. On the other hand, no appreciable change in lattice parameter 'c' has been observed [shown in Fig. 2(b)]. The decrease in 'a' parameter is an indication of the C substitution for B in $MgB_2$ lattice [22,23]. The substituent C atoms are readily available from the C source material i.e. adipic acid. The actual C substitution level for our $MgB_2 + (C_6H_{10}O_4)_x$ samples can be estimated indirectly from the change in lattice parameters using formula $x = 7.5\ \Delta(c/a)$ where $\Delta(c/a)$ is the change in c/a compared to pure sample [3,24]. The net C percentage addition is only 49.32% of the adipic acid ($C_6H_{10}O_4$). Actual C doping level is calculated to be x ~ 0.004 to 0.0295 in the composition of $Mg(B_{1-x}C_x)_2$ as shown in Fig. 2(d) i.e. for the highest doping level of adipic acid the actual C substitution level is 2.95 at% of Boron. For this substitution level of C the decrease in lattice parameter 'a' is in agreement with reported literature [10,25,26]. Fig.2 (c) shows an increase in c/a value as we add the adipic acid in $MgB_2$; which clearly indicates the presence of the lattice strain in doped samples. The lattice strain can be attributed to the C substitution in the structure and unreacted C inside the grains [25]. According to the calculation of single crystal by Lee et al. [27] the actual C substitution level comes out to be 0.0247 for the highest doping level of adipic acid, which is lower than that being calculated from Avdeev et al. [24]. This is discussed in Ref. 3 that according to Avdeev et al the actual carbon content comes out little larger because polycrystalline-carbon-substituted samples may possibly contain some amount of impurity phases with in x-ray resolution limit.

Figure 3 shows the SEM images for (a) undoped and (b) 10 wt% $C_6H_{10}O_4$ doped $MgB_2$. The microstructure of undoped sample appears inhomogenous, consisting of crystalline grains of size from half of micron to one micron. This figure clearly shows the poor connectivity of the undoped sample in comparison to doped sample. In the 10 wt% $C_6H_{10}O_4$ doped $MgB_2$ sample the grains are mixture of platelets and bars and the grain morphology of this sample is refined to smaller and denser compared to undoped sample. The *FWHM* (full width at half maximum) values derived from the XRD pattern (shown in table I) also support the refinement of grains after doping. The small grains are effective in enhancing the flux pinning. In fact the grain boundaries of $MgB_2$ may themselves act as the effective pinning centers [26]. Though the grain size decreases monotonically with an increase in effective C content, the superconducting performance is optimum for 5 wt% $C_6H_{10}O_4$ doped $MgB_2$ sample. This we will discuss later after elaborating upon various superconducting parameters of $C_6H_{10}O_4$ doped $MgB_2$ samples.

Resistivity versus temperature curves for $MgB_2$+(adipic acid)$_x$; (x = 0, 5%, 7% & 10%) samples are shown in Fig. 4. The transition temperature ($T_c$) of pure sample is 38.43 K. The value of $T_c$ for pristine sample is higher than those reported in ref. 10, 25 & 26. Transition temperature decreases continuously with the addition of adipic acid compared to the pristine sample. The $T_c$ of 10 wt% $C_6H_{10}O_4$ doped $MgB_2$ is 34.93 K. The reduction of $T_c$ is due to the increase in C substitution level as we increase the concentration of dopant. The resistivity at 40 K increases from 21 μΩ-cm for the pure $MgB_2$ to 88 μΩ-cm for the 10 wt% $C_6H_{10}O_4$ doped $MgB_2$ compound. The resistivity value at 40 K for our pure $MgB_2$ sample is quite lower than those reported in ref. 28, which is 39.7 μΩ-cm. The higher values of residual resistivity for doped samples indicate that the impurity scattering is stronger due to the carbon substitution at boron sites. The residual resistivity ratio ($RRR = R_{T275K}/R_{Tonset}$) values for the pure, 5 wt% and 10 wt% $C_6H_{10}O_4$ doped samples are 3.06, 1.89 and 1.69 respectively. The *RRR* of pure polycrystalline bulk $MgB_2$ in ref. 27 is 3.0, which is close to the present value and in ref.10 the reported *RRR* value is 2.13, which is lower than our value. After doping in pristine sample the *RRR* comes down, for example *RRR* is only 1.5 for $MgB_{2-x}C_x$ samples [29]. The *RRR* values decreases with increase in doping level (shown in table I). C substitution at B site (revealed by contraction in 'a' parameter and reduction in $T_c$) and the inclusion of unreacted C can enhance the electron scattering, which leads to the monotonic decrease in *RRR* values.

Figure 5 depicts the dc susceptibility versus temperature $\chi(T)$ plots in an applied field of 100 Oe, in field-cooled (*FC*) situations. It is evident from the figure that pure $MgB_2$ undergoes a sharp superconducting transition (diamagnetic, $T_c^{dia}$) at 38.27 K within 1K-temperature interval. All the samples exhibit one-step transition from normal state to superconducting state, however the transition width increases a bit by increasing the amount of dopant. The superconducting critical temperature ($T_c^{dia}$) being seen from $\chi(T)$ measurements is in agreement with the $T_c$ ($\rho = 0$).

The magnetic hysteresis loop for all the doped samples $MgB_2$+(adipic acid)$_x$; (x = 0, 1%, 3%, 5%, 7% & 10%) are shown in Fig. 6(a) at $T$ = 10 K. This figure clearly demonstrates that at $T$ = 10 K the closing of *M(H)* loop for pure sample is at 7.6 Tesla, while the same is closed at 10.4 Tesla for 5 wt% $C_6H_{10}O_4$ doped sample. This indicates that irreversibility field values are improved significantly with addition of adipic acid in parent compound. The variation of irreversibility field with respect to the adipic acid content is shown in Fig. 6(b). The 5 wt% $C_6H_{10}O_4$ doped sample has the highest value of $H_{\mathrm{irr}}$ at all the temperatures. The 7 wt% $C_6H_{10}O_4$ doped sample has also quite competitive values but slightly lower than as for the 5 wt%. The 10 wt% $C_6H_{10}O_4$ doped sample has much lower value of $H_{irr}$. In fact it is lower than that of even the pristine sample. It is clear that the value of $H_{\mathrm{irr}}$ is highest for around 5-7 wt% $C_6H_{10}O_4$ doping in $MgB_2$.

The dependence of critical current density ($J_{\mathrm{c}}$) on the applied magnetic field at 5 and 15 K is shown in Fig. 7 for undoped and 3, 5 & 7 wt% doped samples. The $J_{\mathrm{c}}$ value for the undoped sample is $2.79 \times 10^3$ A/cm$^2$ at 5 K and 8 Tesla whereas for the doped samples the $J_{\mathrm{c}}$ is enhanced to $10^4$ A/cm$^2$ at the same temperature and field. The highest $J_{\mathrm{c}}$ value is achieved for 5 wt% $C_6H_{10}O_4$ doped sample i.e. $2.67 \times 10^4$ A/cm$^2$; which is almost higher by an order of magnitude compared to undoped sample. If we compare this value with reported literature of organic acid doping then we found that this is better than those [10,26]. For example in ref. 10 highest $J_{\mathrm{c}}$ is found to be $1 \times 10^4$ A/cm$^2$ at 5 K in 8 T field; our value is 2.67 times higher than this. Similarly, there is also an increment in $J_{\mathrm{c}}$ value by an order of magnitude at 15 K at higher fields for the 5 wt% $C_6H_{10}O_4$ doped sample with respect to the undoped one; for example the $J_{\mathrm{c}}$ value for undoped sample is $1.6 \times 10^3$ A/cm$^2$ at 6 T & 5 K and the same is increased to $1.56 \times 10^4$ A/cm$^2$ for 5 wt% $C_6H_{10}O_4$ doped sample. As far as the optimization of doping content is concerned, the same is found to be 5wt% addition of adipic acid, which amounts empirically to around 1.37% of C at B-site in $MgB_2$. The enhancement of $J_{\mathrm{c}}$ is attributed to the lattice distortion resulting from the incorporation of C atoms into the $MgB_2$ crystal lattice. Worth mentioning is the fact that besides the lattice distortion, the change of band structure and the changed scattering rates between the two bands are also responsible for the enhancement of critical fields and currents in $MgB_2$ [7]. The introduction of disorder increases scattering of the charge carriers, which reduces their mean free path and improves the upper critical field. As far as the role of grain boundary pinning is concerned, we would like to mention that though the best values of superconducting parameters are achieved for 5 wt% $C_6H_{10}O_4$ doped $MgB_2$ , the grain size is least for 10 wt% $C_6H_{10}O_4$ doped $MgB_2$ sample. It seems that besides the grain boundary pinning effects, some more factors like inhomogeneity, agglomeration and lattice strain etc. also contribute in enhancing the flux pinning strength [25]. Seemingly, in present case, the combination of all the favorable factors is optimum for the 5 wt% $C_6H_{10}O_4$ doped $MgB_2$ sample. In fact the pinning is also expected to be weaker due to lower transition temperature [7] for higher $C_6H_{10}O_4$ doped $MgB_2$ samples. Overall a balanced mechanism is required for getting the best values of superconducting parameters. Grain size alone cannot explain the observed results yet [10,25,26]. In fact besides direct C doping at B-site in $MgB_2$ [14-17], very recently the incorporation of C through various carbohydrates [10,25,26] has attracted a lot of attention. Broadly speaking, it seems that the effect of the most of the used carbohydrates on superconducting performance of $MgB_2$ is more or less the same.

These findings can be further supported by flux pinning results. Fig. 8 shows the dependence of reduced flux pinning force ($F_p / F_{p,\ max}$) of all the doped samples along with undoped one at 15 K. The relationship between flux pinning force and critical current density could be described by [30,31]

$$F_{\mathrm{p}} = \mu_0 \, J_c(H) \, H \qquad (1)$$

Where $\mu_0$ is the magnetic permeability in vacuum. Flux pinning curves for the doped samples are shifted to the right as compared to pure $MgB_2$. This indicates towards the significant improvement in flux pinning forces for adipic acid doped samples in comparison of undoped one i.e. all the doped samples have more pinning centers than the undoped one. Due to the substitution of C at B site the formation of nano-domain structure takes place due to the variation of Mg-B spacing. These nano-domains defects having the size of 2-3 nm can behave as effective pinning centers. Excess carbon, which can be embedded within the grain of $MgB_2$ as nano-inclusions, can also serve as pinning centers.

## Conclusions

In summary, we used adipic acid ($C_6H_{10}O_4$) as the C source material for doping in polycrystalline $MgB_2$. A systematic decrease in lattice parameter 'a' and transition temperature ($T_c$) is observed with increasing amount of dopant. Use of organic acid solution solves the problem of in-homogeneity. Very recently some organic compounds such as sugar [32], benzoic acid [26], mallic acid [25], and tartaric acid [10] have been used as active carbon source for intrinsic pinning in high performance $MgB_2$ superconductor; our results are in confirmation with these reports and on an altogether new additive i.e., adipic acid. It appears to be an inexpensive method instead of using expensive nano-particles [3-6] for improving the superconducting performance of parent $MgB_2$ compound. In addition, the C substitution takes place at the same temperature as the formation of $MgB_2$. The simultaneous activities, substitution of C at B in lattice and the inclusion of excess C within the grains, result in the higher $J_c$ and $H_{irr}$ values for the doped samples. Further we conclude that the impact of various additive organics on superconducting performance of $MgB_2$ superconductor is more or less the same, may it be sugar [32], benzoic acid [26], mallic acid [25], tartaric acid [10] or the presently studied adipic acid.

## Acknowledgement

The authors from *NPL* would like to thank Dr. Vikram Kumar (Director) for his continuous encouragement in present work. Mr. K. N. Sood from *SEM* Division of *NPL* is acknowledged for providing us with the *SEM* micrographs. Arpita Vajpayee would like to thank the *CSIR* for the award of Junior Research Fellowship to pursue the Ph. D. degree.

**Figure & Table captions**

Table I. Critical temperature ($T_c$), residual resistivity ratio (*RRR*), actual carbon substitution level and *FWHM* for the undoped and adipic acid doped $MgB_2$ samples

Figure 1. X-ray diffraction pattern of pure and 1,3,5,7 & 10 wt% adipic acid doped samples

Figure 2. (a) a-axis lattice parameter, (b) c-axis lattice parameter, (c) c/a values and (d) actual amount of C substitution of $MgB_2$+(adipic acid)$_x$ (x = 0, 1%, 3%, 5%, 7% & 10%) samples

Figure 3. SEM pictures of (a) Pure $MgB_2$ & (b)10 wt% adipic acid added sample

Figure 4. Variation of resistivity with temperature $\rho(T)$ plots for Pure, 5, 7 and 10 wt% adipic acid added samples

Figure 5. Magnetization versus temperature plot of $MgB_2$+(adipic acid)$_x$; (x = 0, 1%, 3%, 5%, 7% & 10%) samples

Figure 6. (a) Magnetization loop *M(H)* for $MgB_2$+(adipic acid)$_x$; (x = 0, 1%, 3%, 5%, 7% & 10%) samples up to 13 Tesla field at 10 K and (b) Variation of irreversibility field $H_{irr}$ with respect to adipic acid concentration at 5 K, 10 K & 20 K

Figure 7. $J_c(H)$ plots for adipic acid doped samples along with pristine $MgB_2$ at (a) 15 K & (b) 5 K

Figure 8. Variation of reduced flux pinning force ($F_p/F_{p,max}$) with magnetic field for $MgB_2$+(adipic acid)$_x$; (x = 0, 1%, 3%, 5%, 7% & 10%) at 15 K

Table I.

| $C_6H_{10}O_4$ (wt%) | $T_c$ (K) ($\rho \rightarrow 0$) | $\rho_{40K}$ ($\mu\Omega$-cm) | *RRR* values | Actual C substitution (x) in $MgB_{2-x}C_x$ | *FWHM* of (101) (deg) | *FWHM* of (100) (deg) |
|---|---|---|---|---|---|---|
| 0 | 38.43 | 21.22 | 3.06 | 0 | 0.385 | 0.308 |
| 3 | 36.95 | 107.98 | 2.06 | 0.0108 | 0.492 | 0.372 |
| 5 | 36.62 | 51.69 | 1.89 | 0.0137 | 0.510 | 0.384 |
| 7 | 35.88 | 38.57 | 1.76 | 0.0176 | 0.457 | 0.342 |
| 10 | 34.93 | 88.09 | 1.69 | 0.0295 | 0.514 | 0.373 |

Figure 1.

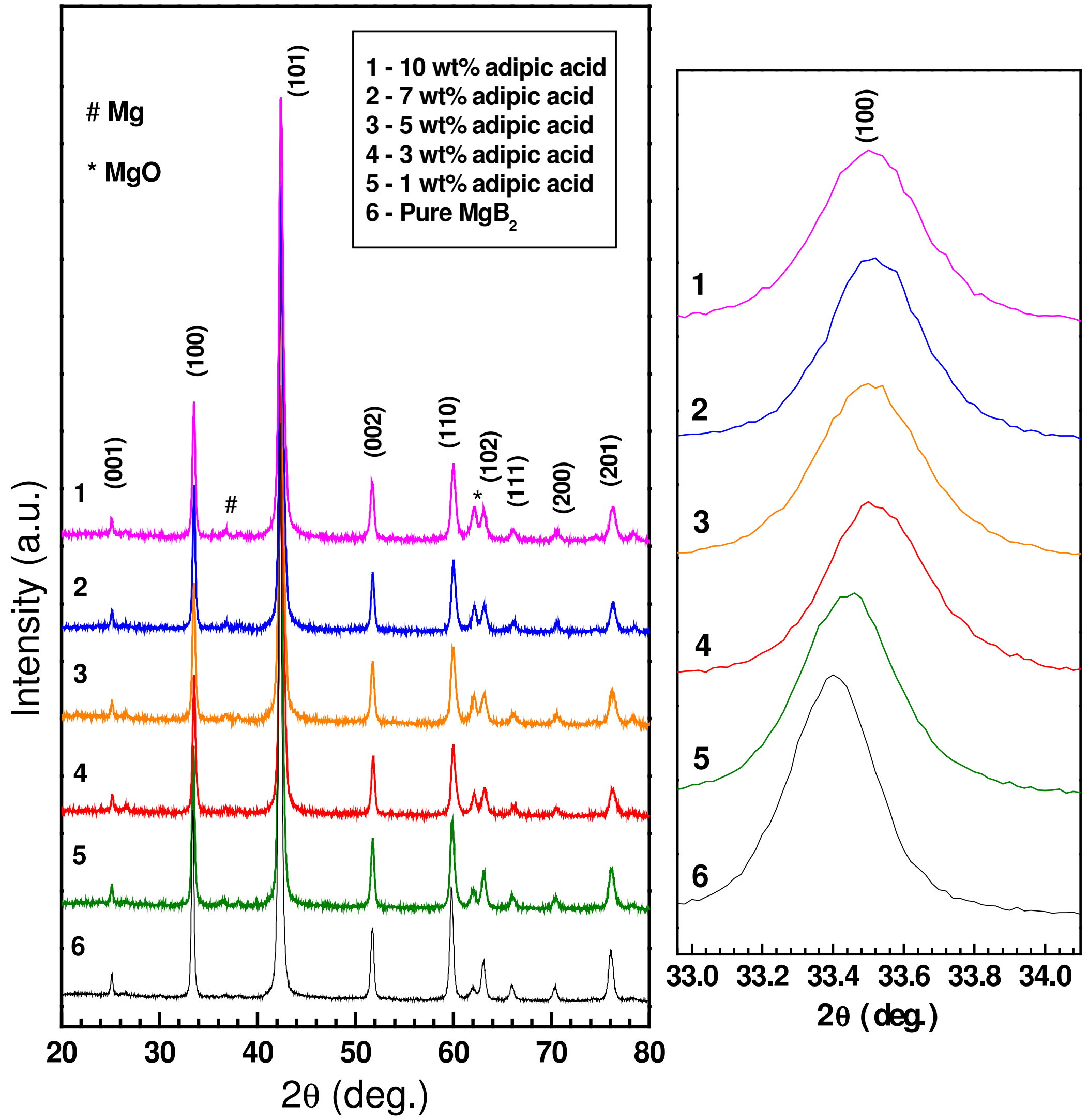

Figure 2.

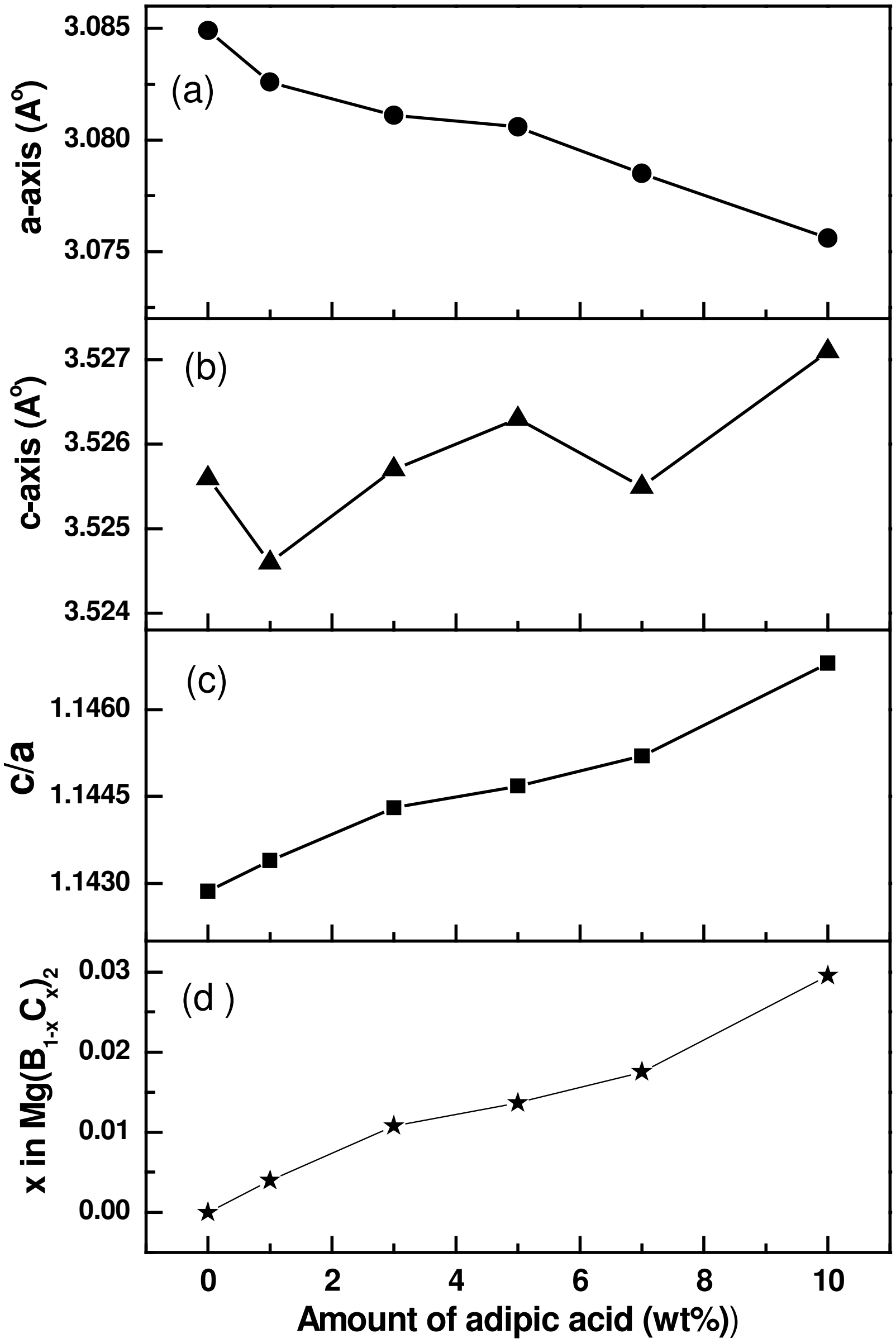

Figure 3.

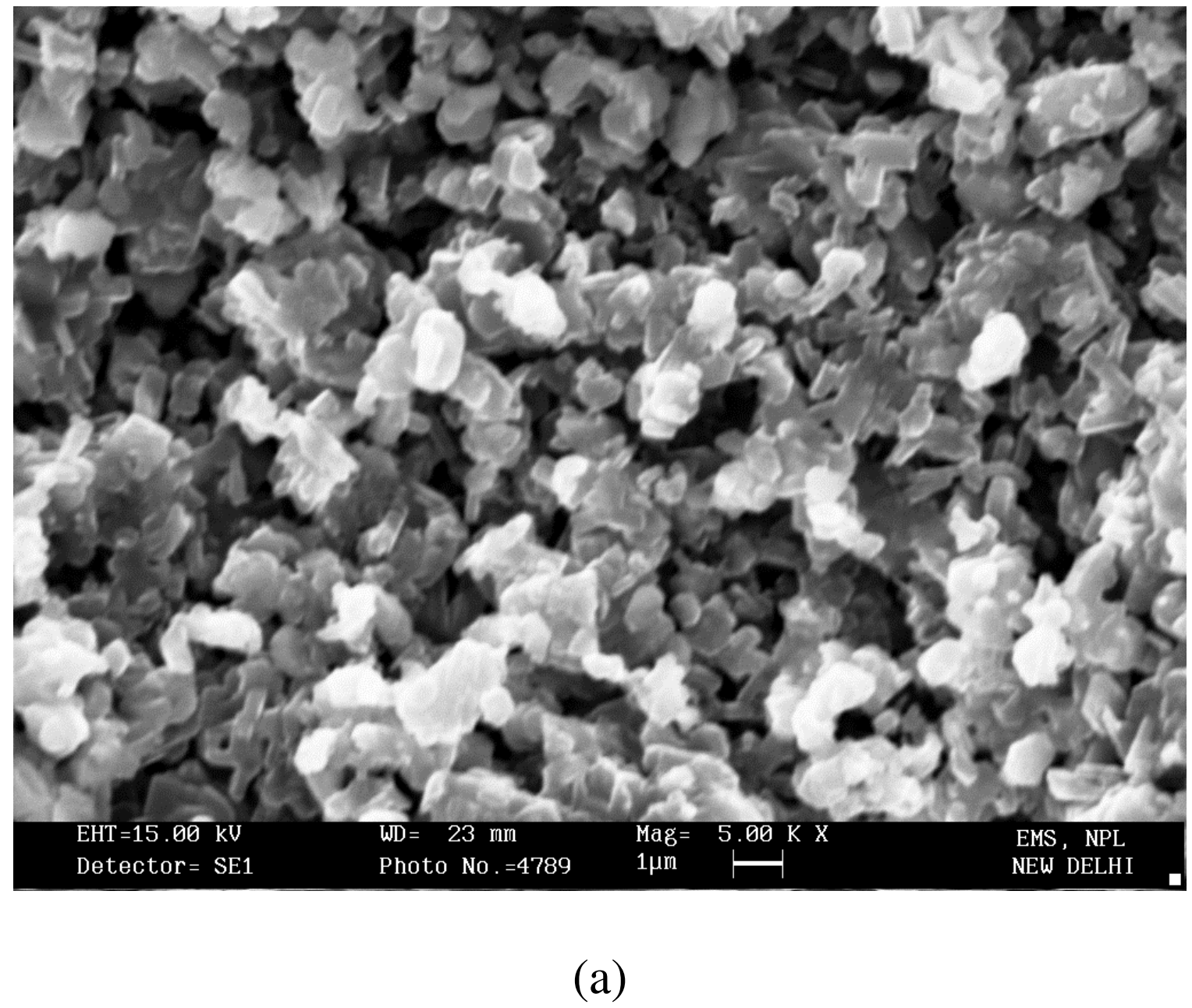

(a)

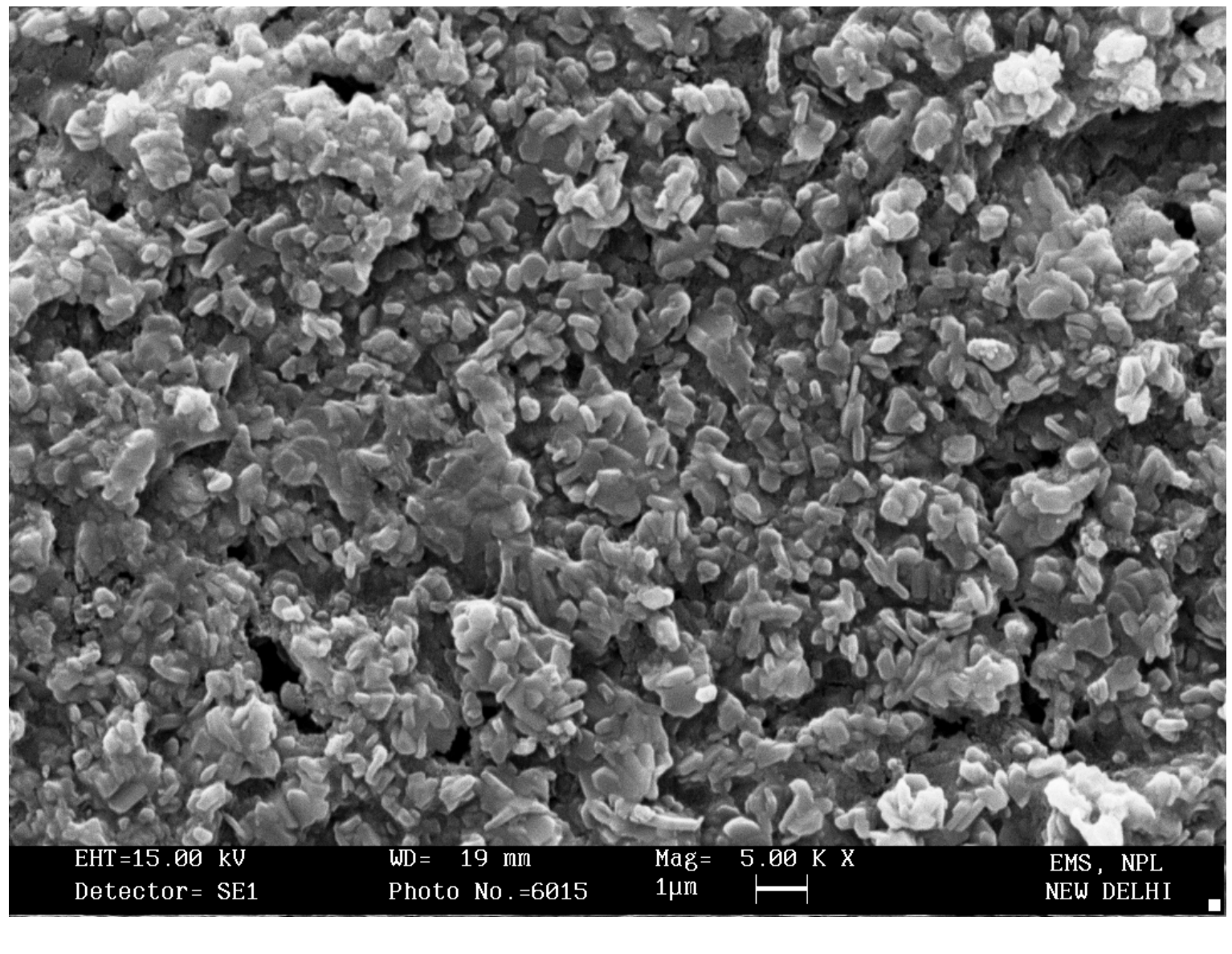

(b)

Figure 4.

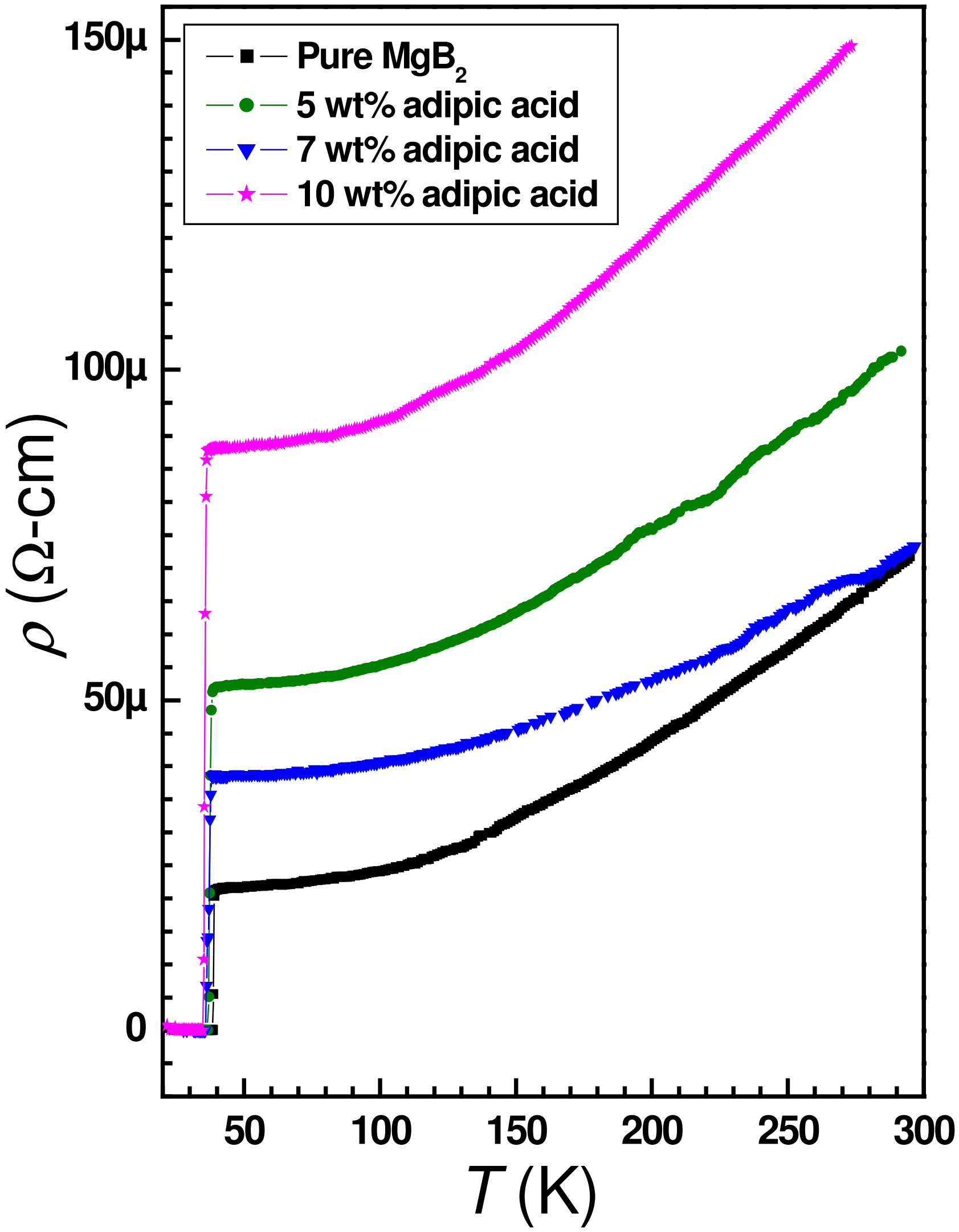

Figure 5.

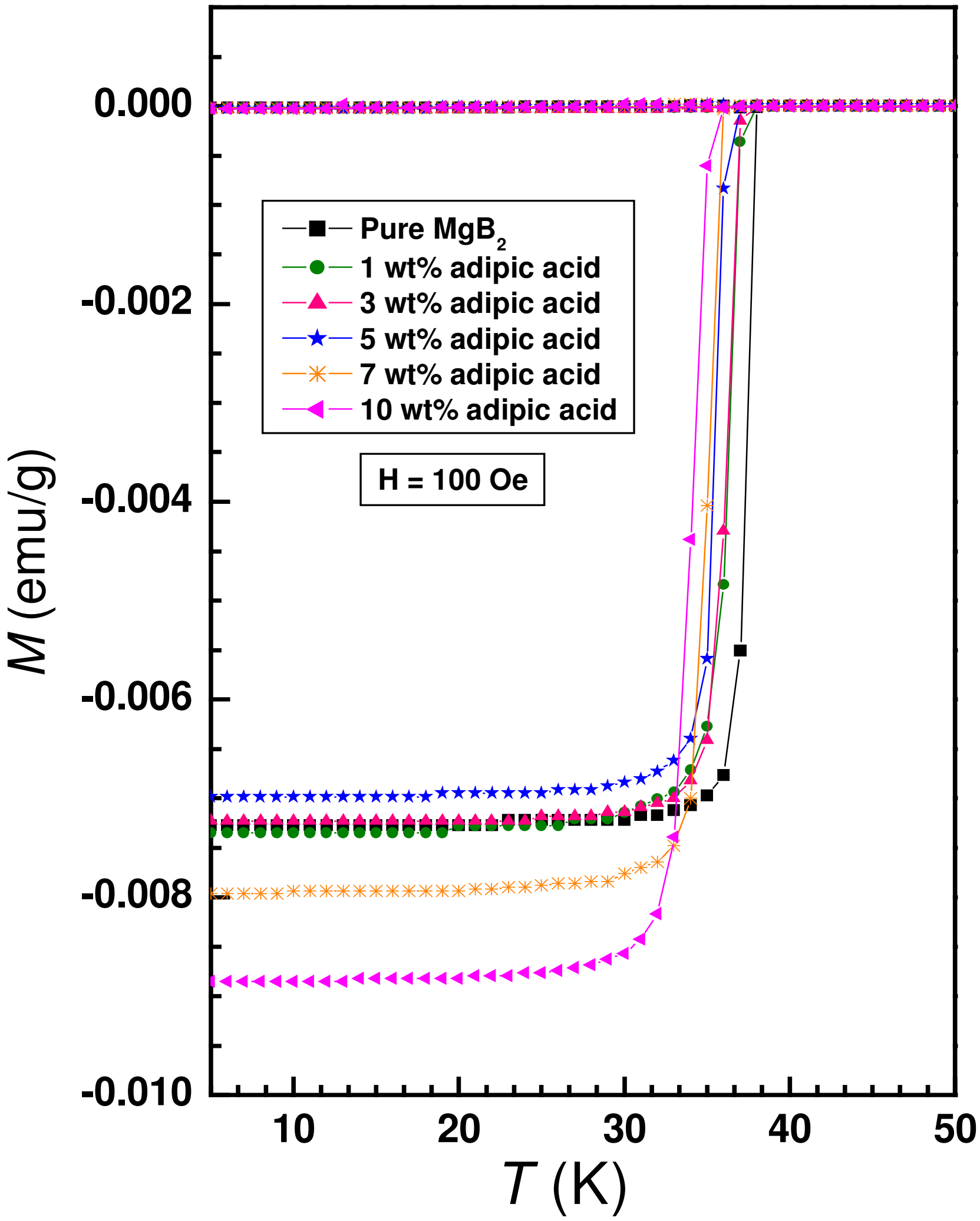

Figure 6 (a).

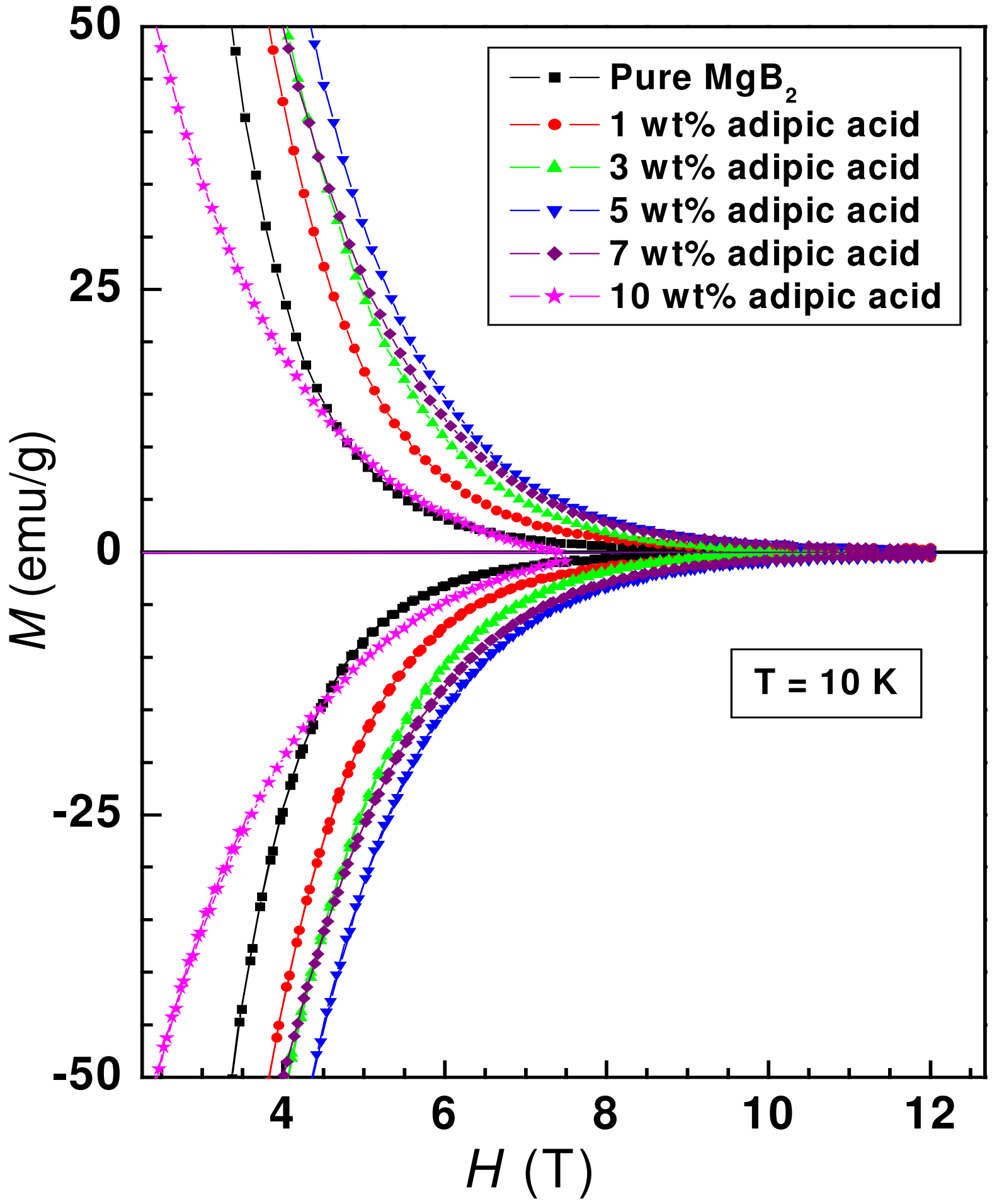

Figure 6 (b).

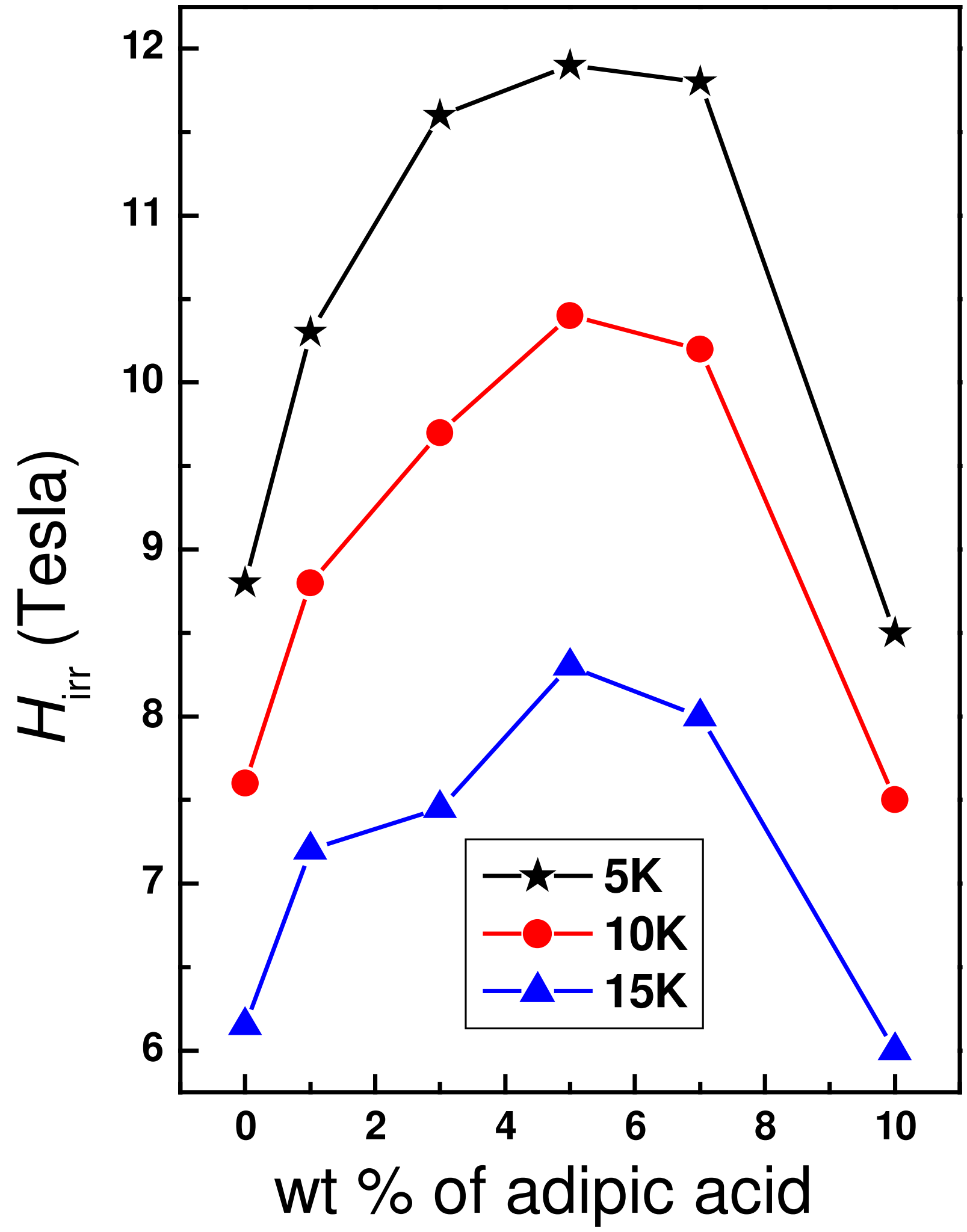

Figure 7

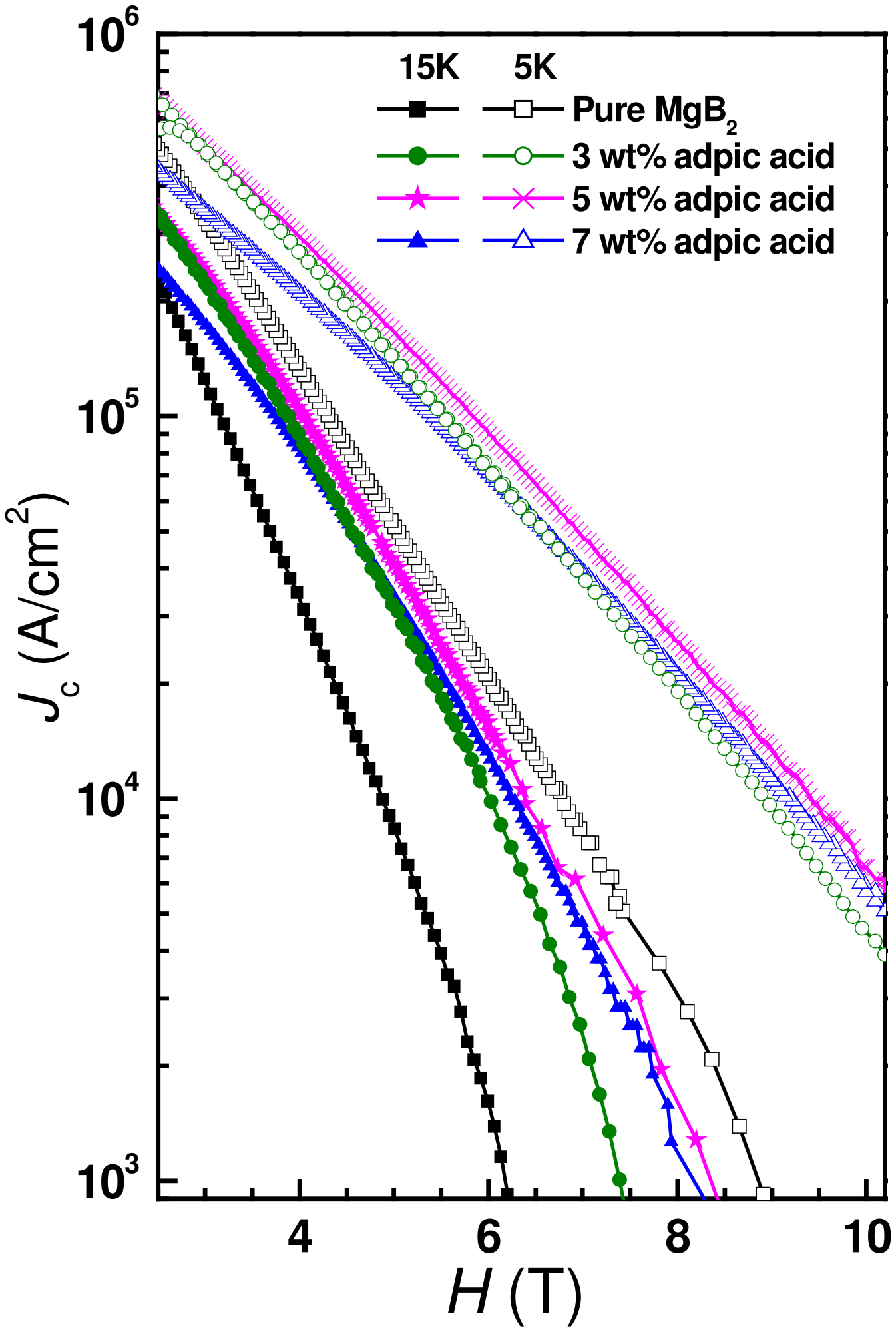

Figure 8.

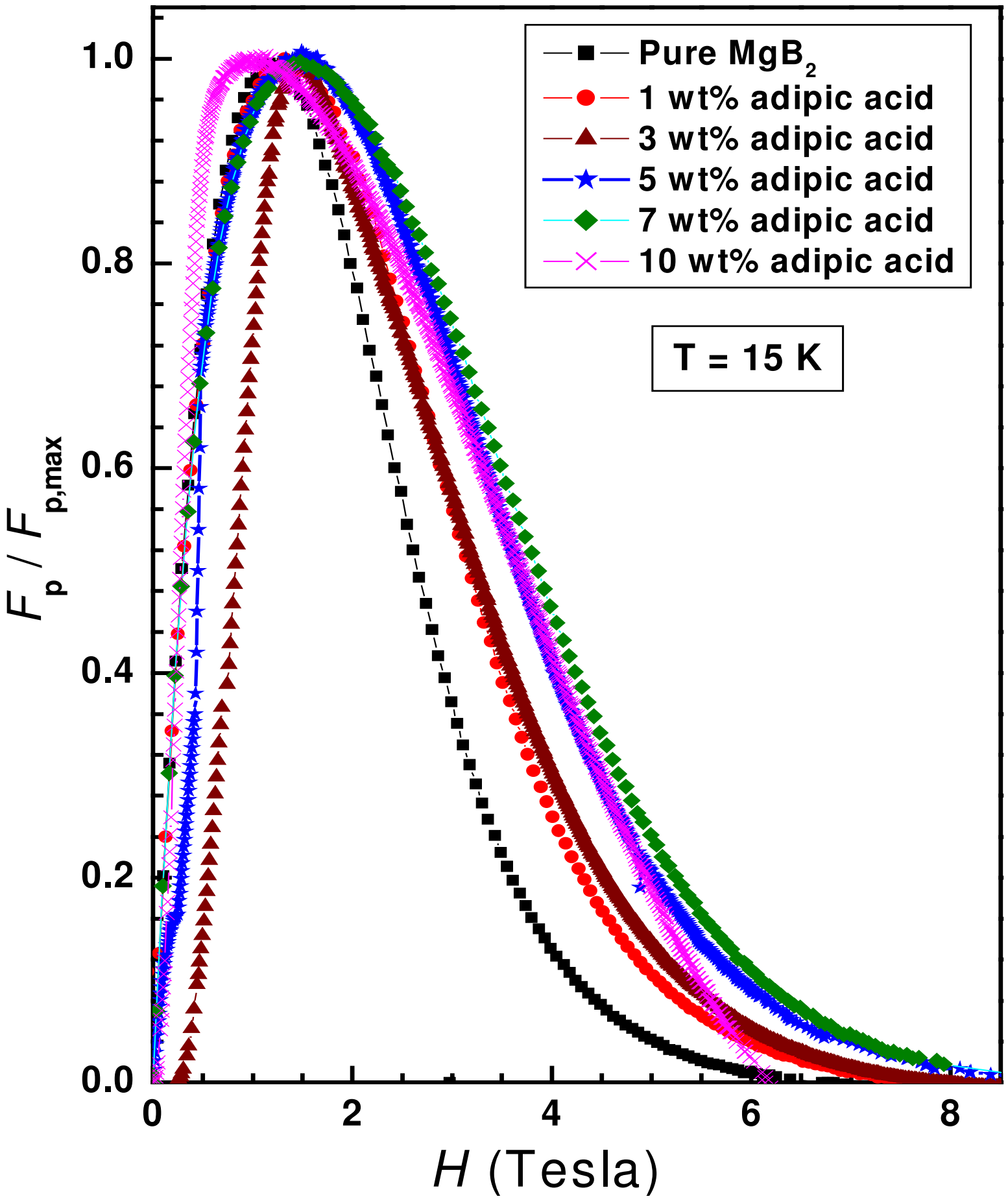